\documentclass[conference]{IEEEtran}
\IEEEoverridecommandlockouts
\usepackage{cite}
\usepackage[table,xcdraw]{xcolor}
\usepackage{amsmath,amssymb,amsfonts}
\usepackage{algorithmic}
\usepackage{multirow}
\usepackage{url}
\usepackage{graphicx}
\usepackage{textcomp}
\usepackage{float}
\usepackage{balance}
\let\oldmathformula\(
\renewcommand{\(}{\oldmathformula\displaystyle}
\usepackage[absolute,overlay]{textpos}
\usepackage{xcolor}
\def\BibTeX{{\rm B\kern-.05em{\sc i\kern-.025em b}\kern-.08em
    T\kern-.1667em\lower.7ex\hbox{E}\kern-.125emX}}
\begin{document}

\title{Automated SAR ADC Sizing Using Analytical Equations}

\author{
    \IEEEauthorblockN{
    Zhongyi Li$^1$\IEEEauthorrefmark{1}, 
    Zhuofu Tao$^2$\IEEEauthorrefmark{1},
    Yanze Zhou$^3$,
    Yichen Shi$^{4,1}$,
    Zhiping Yu$^5$,
    Ting-Jung Lin$^1$\IEEEauthorrefmark{2},
    Lei He$^6$\IEEEauthorrefmark{2}}
$^1$\textit{Ningbo Institute of Digital Twin, Eastern Institute of Technology, Ningbo, China} \\
$^2$\textit{University of California, Los Angeles, USA} \ \ \ $^3$\textit{BTD Technology, Ningbo, China} \\
$^4$\textit{Shanghai Jiao Tong University, Shanghai, China} \ \ \ $^5$\textit{Tsinghua University, Beijing, China} \\
$^6$\textit{Eastern Institute for Advanced Study, Eastern Institute of Technology, Ningbo, China}\\
    \textit{tlin@idt.eitech.edu.cn, lhe@eitech.edu.cn}
    \thanks{This work was partially supported by ``Science and Technology Innovation in Yongjiang 2035" (2024Z283) and by research support from BTD Inc.}
    \thanks{\IEEEauthorrefmark{1} Equal contribution.}
    \thanks{\IEEEauthorrefmark{2} Corresponding authors.}
}

\maketitle

\begin{abstract}


Conventional analog and mixed-signal (AMS) circuit designs heavily rely on manual effort, which is time-consuming and labor-intensive. This paper presents a fully automated design methodology for Successive Approximation Register (SAR) Analog-to-Digital Converters (ADCs) from performance specifications to complete transistor sizing. To tackle the high-dimensional sizing problem, we propose a dual optimization scheme. The system-level optimization iteratively partitions the overall requirements and analytically maps them to subcircuit design specifications, while local optimization loops determines the subcircuits' design parameters. The dependency graph-based framework serializes the simulations for verification, knowledge-based calculations, and transistor sizing optimization in topological order, which eliminates the need for human intervention. We demonstrate the effectiveness of the proposed methodology through two case studies with varying performance specifications, achieving high SNDR and low power consumption while meeting all the specified design constraints.

\end{abstract}

\section{Introduction}

Analog-to-digital converters (ADCs), connecting analog sensors to digital processing, have become an essential IP in most System on Chips (SoCs). Successive Approximation Register (SAR) ADCs, with applications ranging from medium-low precision to ultra-high-speed designs, have attracted significant attention \cite{budak2021dnn, tang2022low, murmann2016successive}. However, conventional design methods primarily rely on manual adjustments. As technology nodes shrink and design complexity grows exponentially, manually designing SAR ADCs has become increasingly time-consuming. This inefficient design process struggles to meet the demands of a fast-evolving market, necessitating an automatic design methodology. Recently, there has been many data-driven works for automated circuit design~\cite{tao2024amsnet, shi2024amsnetkg}, this motivates us to develop a similar approach.

Device sizing is a crucial step in the design flow of analog and mixed-signal (AMS) integrated circuits, such as SAR ADC. Previous work on automatic sizing can be broadly classified into knowledge-based and optimization-based approaches. Knowledge-based approaches rely on analytic equations to derive circuit performance and device sizing, usually involving assumptions and approximations with various empirical values. In contrast, optimization-based approaches depend on trials and errors in the simulation environments and are very time-consuming. Significant research effort \cite{budak2021dnn, xing2024kato, li2024high, mina2022review} has been dedicated to automating optimization-based methods using machine learning (ML) algorithms such as Bayesian optimization (BO) and reinforcement learning (RL). However, these methods are often limited to small-scale circuits, such as OPAMP or comparator designs within twenty transistors. Due to the extensive parameter search spaces and high simulation costs, only a few studies \cite{huang2015systematic, ding2018hybrid, liu2021opensar} have been extended to larger ones, such as ADCs. On the other hand, there have been some pioneering knowledge-based automatic sizing \cite{liu2024ladac}, but again the circuit scale is limited.

\begin{figure}[t]
    \centering
    \includegraphics[width=0.8\columnwidth]{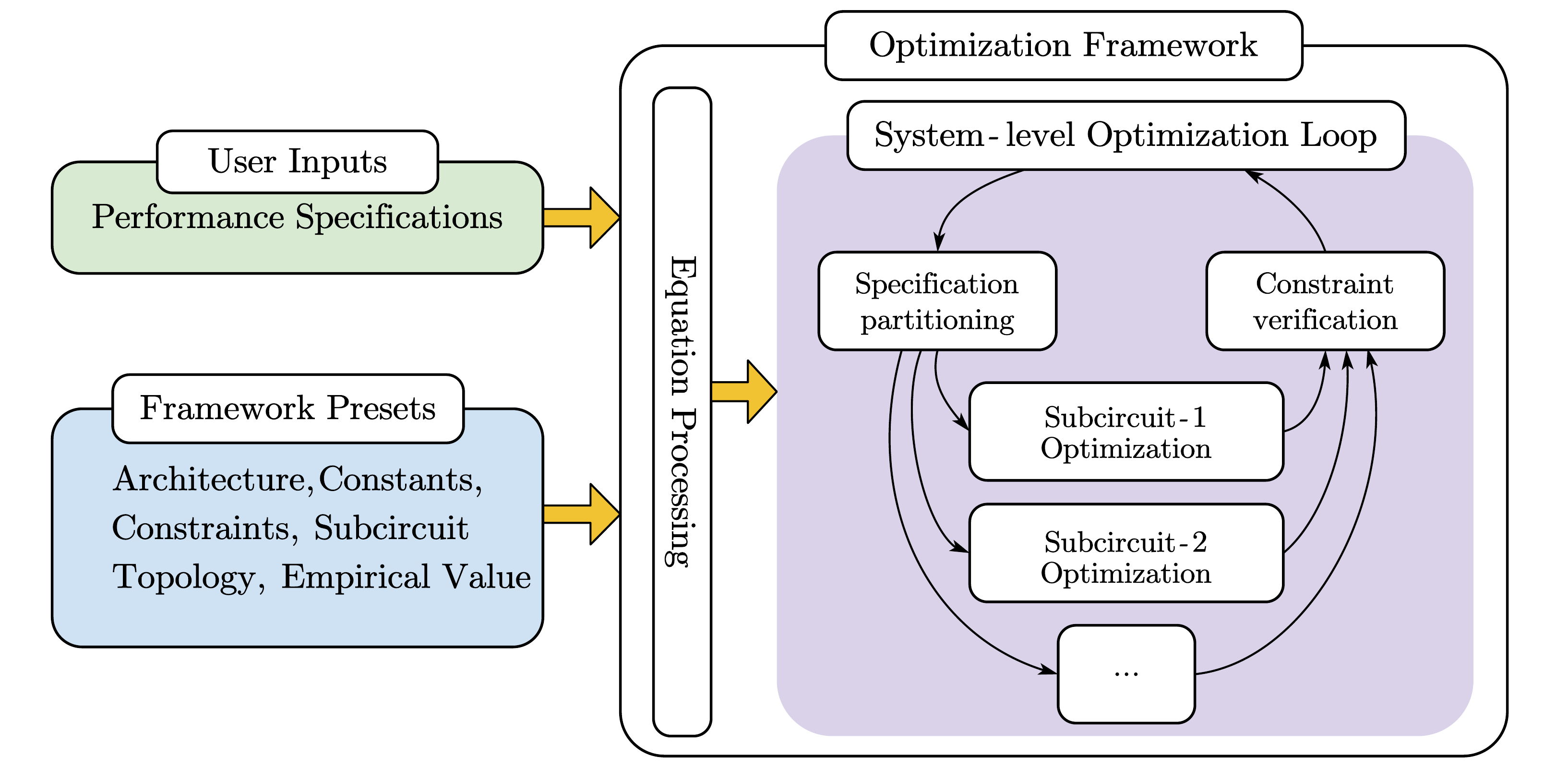}
    \caption{The Proposed Automated Design Framework}
    \label{fig:overview}
\end{figure}

There have been research achievements on automatic sizing of SAR ADCs. However, the proposed methods typically require human interventions within the optimization loop. This demands professional SAR-ADC-specific knowledge, and could potentially introduce human errors. \cite{huang2015systematic, ding2018hybrid} both achieve some level of automation in SAR ADC design. However, \cite{huang2015systematic} still requires a human expert to determine the priority of device sizing. \cite{ding2018hybrid} needs a pre-characterized comparator library of different device parameters, which is time-consuming to obtain.

To address the above challenges, this paper proposes an automated system-level design methodology for SAR ADCs, as illustrated in Fig.~\ref{fig:overview}. The overall motivation is to make use of numerous analytical equations that maps performance from complex circuits to its simpler subcircuits, where latter are usually within reach of popular automated sizing frameworks. The tool requires only high-level performance specifications as input, and the built-in library will supply the circuit topology, testbenches, and analytical equations, facilitating automatic generation of complete SAR ADC netlists without human intervention. It begins with an equation processing phase, during which a dependency graph is constructed from the analytical equations to establish a topological order for simulations for verification, knowledge-based calculations and optimization-based transistor sizing. Based on the dependency graph, we integrate a dual optimization scheme to address the high-dimensional transistor sizing optimization problem in SAR ADCs. The system-level optimization iteratively decomposes the overall performance specifications into individual subcircuits, while several local optimization loops employ BO to optimize the transistor parameters of the subcircuits. The locally optimized results of the subcircuits are then fed back to the system level for overall performance simulation or calculation. Finally, the loop concludes once all constraints are satisfied, yielding the complete device sizing for the SAR ADC.

Our contributions are as follows.
\begin{itemize}
    \item We propose a fully-automated SAR ADC design methodology that generates optimized device sizing directly from performance specifications. The dependency graph-based framework is proposed to serialize the simulations for verification, knowledge-based calculations, and transistor sizing optimization in topological order.
    \item We utilize a dual optimization scheme to tackle the high-dimensional sizing problem. The system-level optimization iteratively decomposes the overall performance specifications into individual subcircuits, while local optimization loops employ BO to optimize the transistor parameters of the subcircuits. This ensures that both overall performance goals and individual subcircuit constraints are met efficiently.
    \item We demonstrate the effectiveness of our framework through two case studies with varying performance specifications.
\end{itemize}

\section{Preliminaries}
\label{sec:prelimanaries}

This section introduces the background knowledge of SAR ADCs including the key design considerations. We also provide a brief overview of Bayesian optimization (BO), the machine learning algorithm employed to optimize device parameters and empirical values.

\subsection{SAR ADC Architecture}

\begin{figure}[t]
    \centering
    \includegraphics[width=\columnwidth]{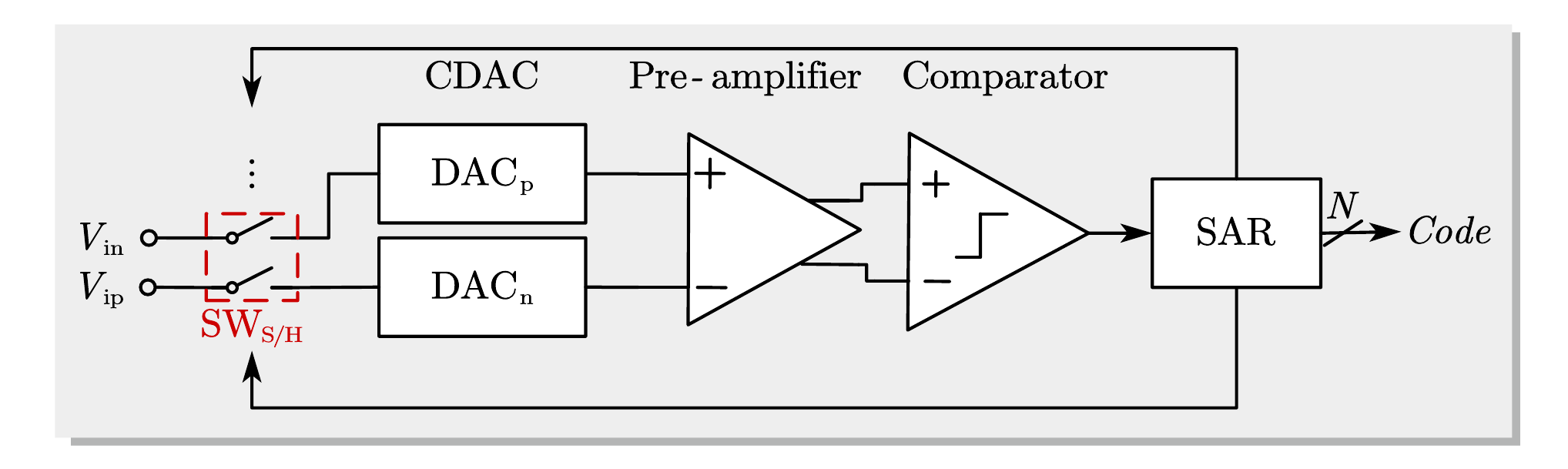}
    \caption{A typical SAR ADC architecture, which is assumed in this work}.
    \label{fig:architecture}
\end{figure}

Fig. \ref{fig:architecture} shows a typical SAR ADC architecture. The functionality of each subcircuit is listed below.

\begin{itemize}
    \item \textbf{Bootstrap Sample-and-Hold Switch (SW$_{\textbf{S/H}}$):} Captures and stores the input voltage during the sampling phase to provide a stable input signal for conversion.
    \item \textbf{Capacitive Digital-to-Analog Converter (CDAC):} Converts the digital code from the SAR logic into an analog voltage for comparison.
    \item \textbf{Pre-amplifier:} Amplifies the input signal to a desired level, enhancing accuracy and minimizing noise before comparison.
    \item \textbf{Comparator:} Compares the CDAC outputs and generates a digital decision.
    \item \textbf{SAR Logic:} Controls the iterative conversion process, adjusting the CDAC and comparator settings to produce the final digital output.
\end{itemize}

The main performance metrics of the SAR ADC usually include resolution \(N\), sampling frequency $f_s$, power consumption, and area.

\subsection{Bayesian Optimization}
Eq. \ref{eq:opt_problem} describes an abstract optimization problem, where we seek to maximize a certain objective function $F(\mathbf{x})$ by tuning the input variables $\mathbf{x}$, subject to a series of $p$ constraints $g_j(\mathbf{x}) \le 0$. In the context of device sizing, $F(\mathbf{x})$ is typically defined by some performance metrics, where each $g_j(\mathbf{x})$ describes the legal search space of a certain parameter, usually set by the PDK or prior knowledge.
\begin{equation}
\label{eq:opt_problem}
\begin{aligned}
& \text{maximize} \quad F(\mathbf{x}) \\
\text{s.t.} \quad & g_j(\mathbf{x}) \leq 0, \quad \forall j \in \{1, \ldots, p\} \\
\end{aligned}
\end{equation}

Bayesian optimization uses a machine learning model, typically a Gaussian process, as the surrogate model to represent the true relationship between the input parameters and the output objective function. The optimization process is iterative; in each iteration, it interacts with the unknown function $F(\mathbf{x})$ with some input $\mathbf{x}$ carefully chosen by its acquisition function, and uses the output to update the surrogate model. After sufficient iterations, it is expected to converge to a sufficiently strong objective.

\section{Algorithms}
\label{sec:method}

\begin{table*}[htbp]
\caption{Variables and Relations}
\renewcommand{\arraystretch}{1.6}
\resizebox{\textwidth}{!}{%
\begin{tabular}{|c|c|c|}
\hline
 & \textbf{Variables} & \textbf{Relations} \\
\hline \hline
\textbf{System-Level} & \textbf{N}, $\mathbf{F_s}$, $\mathbf{V_{\textbf{fs}}}$, $\mathbf{V_{\textbf{DD}}}$ $\mathbf{R_{\textbf{on,TG}}}$ & $\mathbf{V_{\textbf{DD}}} = 0.9$, $\mathbf{R_{\textbf{on,TG}}} = 150$ \\
\hline
\textbf{Empirical Values} & \textbf{D}, \textbf{E}, \textbf{PSA}, \textbf{PSTR} & $1 \le$ \textbf{D} $\le 3$, $1 \le$ \textbf{E} $\le 3$, $0.5 \le$ \textbf{PSA} $\le 0.9$, $0.5 \le$ \textbf{PSTR} $\le 0.9$ \\
\hline
\textbf{Bootstrap} & \textbf{ENOB}, $\mathbf{C_{\textbf{L}}}$, $\mathbf{R_{\textbf{on}}}$, $\mathbf{R_{\textbf{on},\textbf{max}}}$ & 
\begin{tabular}{c}
\textbf{ENOB} $\ge N + D$, \quad $\mathbf{C_{\textbf{L}}} \geqslant \dfrac{6 \cdot 2^{2N-2} kT}{V_{\text{fs}}^2\left( 10^{\frac{1}{10}}-1 \right)}$ \\ 
$\mathbf{R_{\textbf{on},\textbf{max}}} = \max\left\{ R_{\text{on,TG}}, R_{\text{on,BS}} \right\}$
\end{tabular} \\
\hline
\textbf{CDAC} & $\mathbf{C_u}$, $\mathbf{\sigma_u}$, $\mathbf{\sigma_{u,\textbf{constraint}}}$, $\mathbf{C_{\textbf{L},\textbf{max}}}$ & 
\begin{tabular}{c}
$\mathbf{C_u} = \dfrac{C_L}{2^{N/2}}, \quad \mathbf{\sigma_u} \leqslant \sigma_{u,\mathrm{constraint}}$, \quad $\mathbf{\sigma_{u,\textbf{constraint}}} = \dfrac{C_u}{6\sqrt{2^N-1}}$ \\ 
$\mathbf{C_{\textbf{L},\textbf{max}}} = \left( 2^{N/2}-1 \right) C_u$
\end{tabular} \\
\hline
\textbf{Preamplifier} & $\mathbf{A_V}$, $\mathbf{f_{-3\textbf{dB}}}$ & 
\begin{tabular}{c}
$\mathbf{T_\textbf{comp,constraint}} = 1.2R_{\text{on},\max}C_{L,\max} \ln \left( 2^{N+E} \right)$, \quad $\mathbf{A_V} \geqslant 20\log_{10}\left( V_{\text{os}} 2^{N+1} \right)$ \\ 
$\mathbf{f_{-3\textbf{dB}}} \geqslant \dfrac{\ln \left( 1 - \text{PSA} \right)}{-2 \pi \text{PSTR}\cdot T_{\text{comp}}}$
\end{tabular} \\
\hline
\textbf{Comparator} & $\mathbf{T_\textbf{comp,constraint}}$, $\mathbf{T_\textbf{pd}}$, $\mathbf{V_{\textbf{os}}}$, $\mathbf{T_\textbf{comp}}$ & $\mathbf{T_{\textbf{pd}}} \leqslant 0.2 T_{\text{comp},\text{constraint}}$, $\mathbf{T_\textbf{comp}} = \frac{1}{f_s}\frac{1}{\left( N+n \right)} \geqslant T_{\text{comp},\text{constraint}}$ \\
\hline
\end{tabular}
}
\label{tab:variables_and_relations}
\end{table*}

\begin{figure*}[htbp]
\centering
\includegraphics[width=\linewidth]{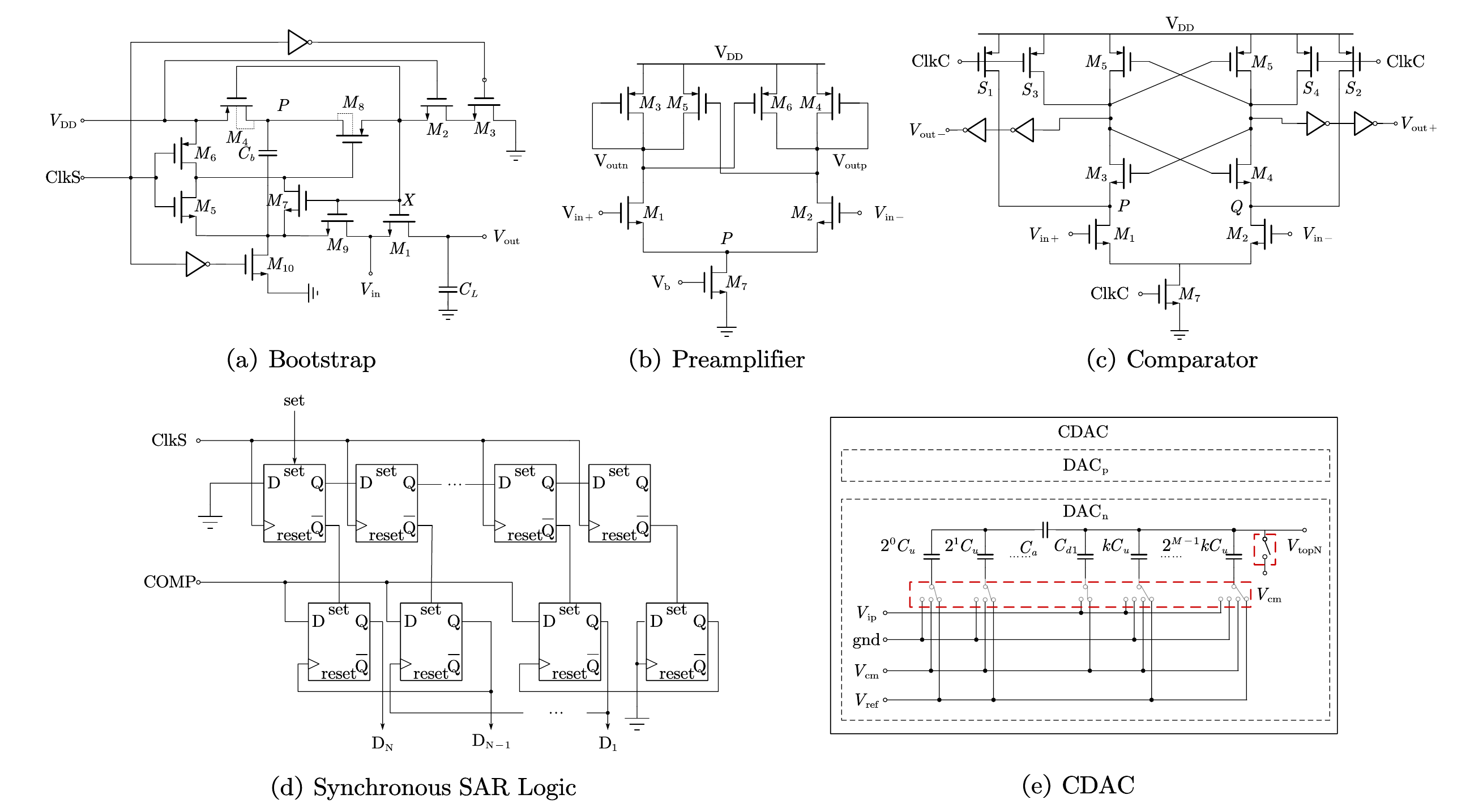}
\caption{Circuit topologies of SAR ADC subcircuits}
\label{fig:Block Diagram of SAR ADC Subcircuits}
\end{figure*}

\begin{figure*}[htbp]
\centering
\includegraphics[width=\linewidth]{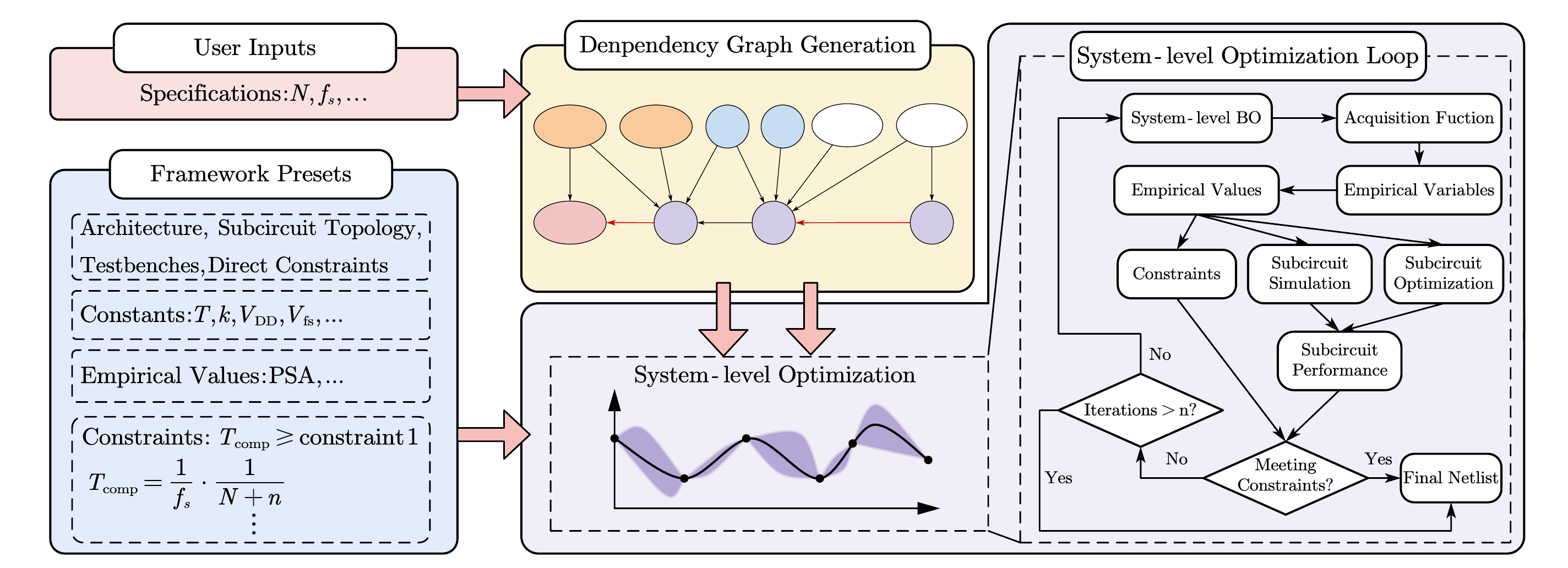}
\caption{The flowchart of the proposed automated SAR ADC design methodology}
\label{fig:Overall_design_flow}
\end{figure*}

In this section, we introduce the algorithms in the proposed optimization framework shown in Fig.  \ref{fig:Overall_design_flow}. 
Due to its complexity, device sizing for SAR ADCs cannot be achieved holistically. A divide-and-conquer design approach is required to decompose the problem into subcircuit-level challenges, making them easier to address. This is accomplished through metric mapping, where the performance specifications of the complete design are analytically translated into the specifications for each individual subcircuit. Each subcircuit is then optimized independently and, once finalized, integrated into the full SAR ADC for performance verification.

Section \ref{sec:method_subcircuits} first formulates the overall and subcircuit-level requirements as \textbf{variables} and \textbf{relations}. Section \ref{sec:method_dependency_graph} presents the \textbf{dependency graph} used to first partition specifications to subcircuits, then serialize the analytical calculation, subcircuit optimization, and constraint verification steps. Section \ref{sec:method_bayesian_optimization} describes how BO is applied in two distinct contexts: \textbf{subcircuit-level} BO to optimize individual subcircuits and \textbf{system-level} BO to optimize the overall design. 



\subsection{Subcircuit Design Specifications}
\label{sec:method_subcircuits}

In the architecture shown in Fig. \ref{fig:architecture}, the subcircuits we assume are presented in Fig. \ref{fig:Block Diagram of SAR ADC Subcircuits}. The optimization objectives and design constraints are summarized in Table \ref{tab:variables_and_relations}. 
Subcircuit performance metrics are constrained not only by overall performance goals but also by interdependencies among them. For clarity, we define a \textbf{variable} as any constant (e.g., the Boltzmann constant $k$), empirical value (e.g., selecting an \textit{opamp setup accuracy} between 0.5 and 0.9), overall performance goal (e.g., ADC resolution $N$), or subcircuit performance goal (e.g., preamplifier gain). Similarly, we define a \textbf{relation} as any equation or inequality constraint. With this in mind, design considerations of subcircuits are presented as follows.

\textbf{Bootstrap sampling switch:} Fig. \ref{fig:Block Diagram of SAR ADC Subcircuits}(a) presents the bootstrap circuit. The load capacitance \(C_L\) is determined by the noise constraint given by 
\begin{equation}
C_L \geqslant \frac{6 \cdot 2^{2N - 2} \cdot kT}{V_{\text{fs}}^2 \left( 10^{\frac{1}{10}} - 1 \right)}
\end{equation}
where \(N\) is the ADC resolution, \(k\) is the Boltzmann constant, \(T\) is the absolute temperature, and \(V_{\text{fs}}\) is the input amplitude.

The effective number of bits (ENOB) is an optimization objective which must satisfy the following constraint
\begin{equation}
\mathrm{ENOB} \geqslant \mathrm{N} + \mathrm{D}
\end{equation}
where \(\mathrm{D}\) is an empirical margin to ensure that the sampling accuracy of the bootstrap switch meets design requirements.

\textbf{CDAC:} Fig.~\ref{fig:Block Diagram of SAR ADC Subcircuits}(e) presents the segmented capacitive array of the CDAC. The unit capacitance and standard deviation are denoted by $C_u$ and $\sigma_u$, respectively, with the following constraint:
\begin{equation}
    \frac{\sigma_u}{C_u} < \frac{1}{6 \sqrt{2^N - 1}}
\end{equation}

\begin{figure}[H]
    \centering
    \includegraphics[width=\columnwidth]{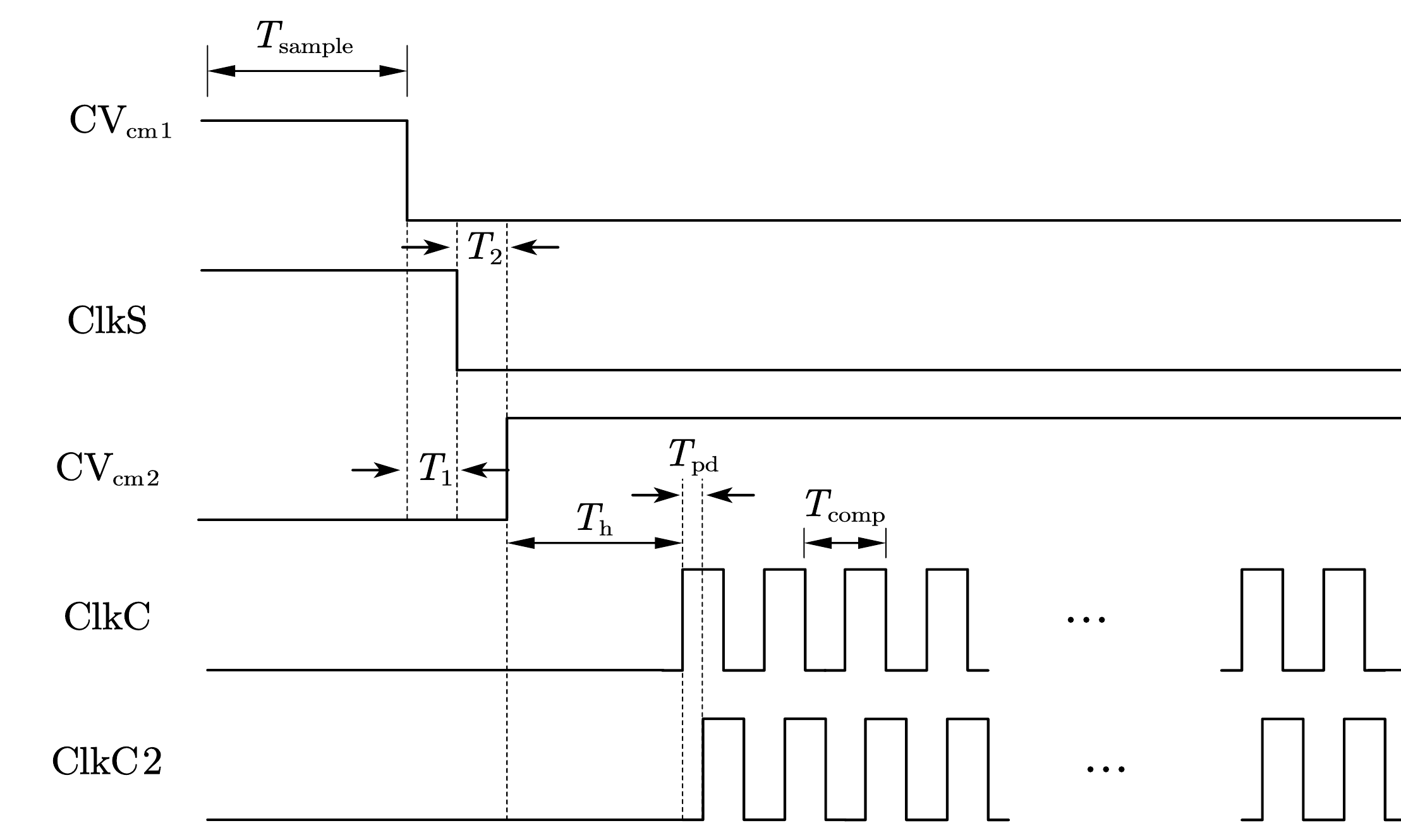}
    \caption{Timing diagram of the clock signals}
    \label{fig:Clock Signal}
\end{figure}

\textbf{Preamplifier:} The preamplifier is a three-stage cascaded structure, where a single stage is shown in Fig. \ref{fig:Block Diagram of SAR ADC Subcircuits}(b). The required gain \(A_V\) of the preamplifier is determined by the comparator offset voltage $V_{\text{os}}$ and $N$, as shown below.
\begin{equation}
A_V \geqslant 20\log _{10}\left( V_{os} 2^{N+1} \right)
\end{equation}
Additionally, the preamplifier bandwidth \(f_\text{-3dB}\) is constrained by the comparator comparison time $T_{\text{comp}}$, as shown in Fig. \ref{fig:Clock Signal}, where the preamplifier needs to amplify the input signal within a portion of $T_{\text{comp}}$. We define this portion as the preamplifier setup time ratio (\(\mathrm{PSTA}\)), expressed as a fraction of the comparator clock cycle, with a range of [0.5, 0.9]. On the other hand, the preamplifier setup accuracy (\(\mathrm{PSA}\)) refers to the accuracy of the preamplifier’s output value within the setup period, with a range of [0.5, 0.9]. The constraint works out to be:
\begin{equation}
f_{-3\mathrm{dB}} \geqslant \frac{\ln \left( 1 - \mathrm{PSA} \right)}{-2 \cdot \pi \cdot \mathrm{PSTR} \cdot T_{\text{comp}}}
\end{equation}

\begin{figure}[H]
    \centering
    \includegraphics[width=\columnwidth]{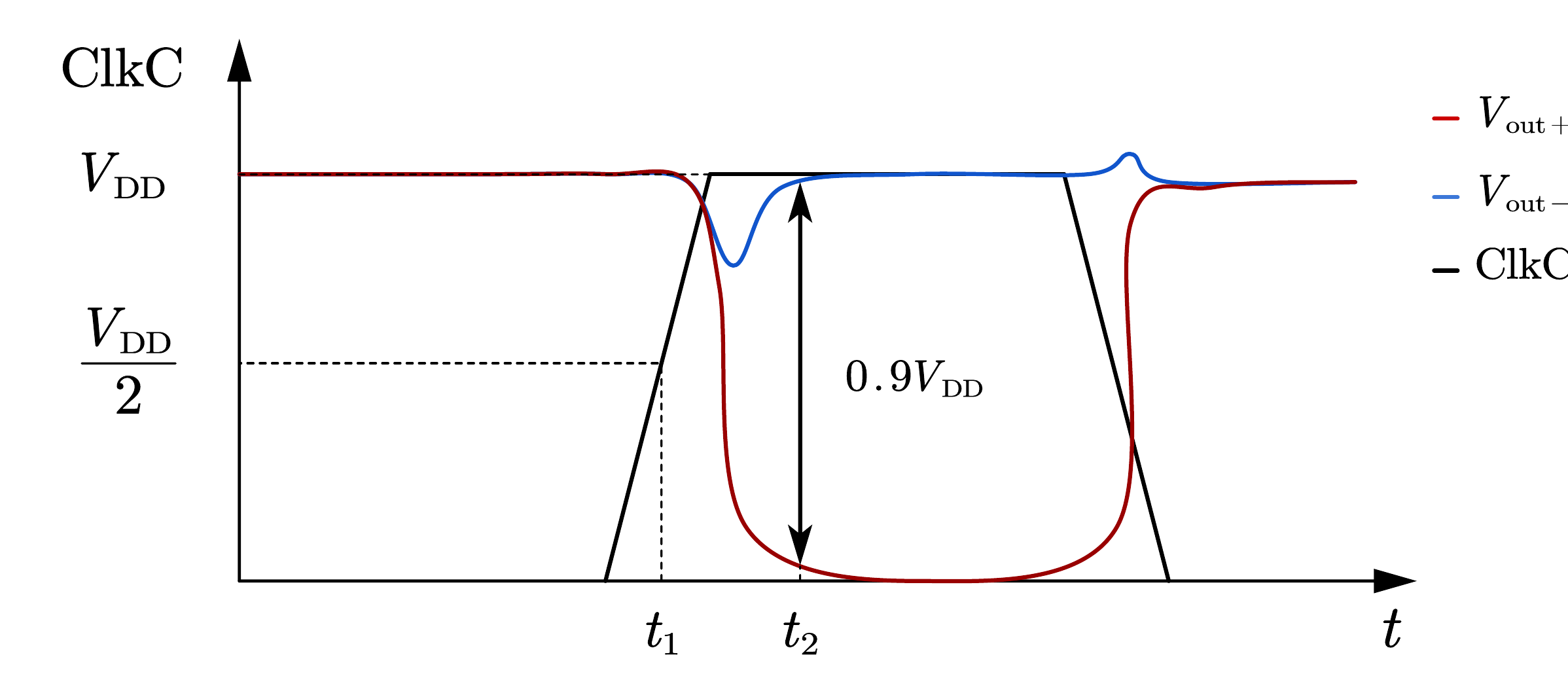}
    \caption{Definition of $T_\text{pd}$: comparator propagation delay \((t2-t1)\)}
    \label{fig:Comparator Speed}
\end{figure}

\textbf{Comparator:} Fig. \ref{fig:Block Diagram of SAR ADC Subcircuits}(c) is a StrongARM latch comparator widely used in SAR ADCs due to its high speed, accuracy, and noise immunity.
The main performance metrics for the comparator design are the comparison period \(T_{\text{comp}}\), comparator propagation delay \(T_{\text{pd}}\), and the offset voltage \(V_{os}\).
The propagation delay $T_{\text{pd}}$ is shown in Fig. \ref{fig:Comparator Speed}, which represents the time interval from the start of the comparator's comparison to the start of the SAR logic operation. The comparison must complete within this time.

For the comparator offset \(V_{os}\), to improve the accuracy of the comparator, a preamplifier is typically added, which amplifies the input signal, allowing the comparator to more accurately detect the difference between the input signal.






\textbf{SAR logic and clock control:} Presented in Fig.~\ref{fig:Block Diagram of SAR ADC Subcircuits}(d), SAR logic and clock control are crucial for SAR ADC performance, impacting accuracy, speed, power consumption, and stability. SAR logic controls the successive approximation process, while clock control ensures precise timing for sampling and comparison, making optimization essential for high-efficiency, high-accuracy ADCs.

As shown in the Fig. \ref{fig:Clock Signal}, \(T_\text{sample}\) represents the sampling time, \(T_1\) and \(T_2\) are the timing for the bottom-plate sampling, which helps reduce signal-dependent charge injection. \(T_\text{h}\) represents the hold time, and \(T_\text{comp}\) is the comparison period of the comparator. In this paper, we set the sampling time \(T_\text{sample}\) equal to one comparison period \(T_\text{comp}\).

For the sampling time and hold time, it is necessary to ensure that the CDAC settles with sufficient accuracy during this period. For the hold time, since bottom-plate sampling is used, the CDAC’s settling time and the preamplifier’s transfer time must be considered. For the hold time and the comparator’s comparison period, which is an integer multiple of the entire quantization cycle, it is required to ensure that the CDAC completes the settling stage during the time interval \(T_\text{comp}- T_{pd}\) after the comparison.

For the CDAC, other than the input signal, all other signals are controlled by transmission gate switches. When the switches are on, the switch and CDAC can be viewed as a low-pass filter, thus the constraints are:
\begin{flalign}
        T_{\text{comp}} &= \frac{1}{f_s} \times \frac{1}{N + n} \geqslant 1.2 R_{\text{on}, \max} C_{L, \max} \ln \left( 2^{N + E} \right) && \\
        T_h &\geqslant 1.5 t_0 = 1.5 R_{\text{on}, \max} C_{\text{tot}} \ln \left( 2^{N + E} \right) && \\
        T_{\text{pd}} &\leqslant \frac{T_{\text{comp}}}{5} && 
\end{flalign}

where \(R_{\text{on}, \max} = \max \left\{ R_{\text{on}, BS}, R_{\text{on}, TG} \right\}\), with \(R_{\text{on}, BS}\) being the on-resistance of the bootstrap sampling switch, and \(R_{\text{on}, TG}\) being the on-resistance of the transmission gate switch. \(C_{L, \max}\) is the capacitance of the MSB capacitor, representing the maximum load capacitance connected to the switch during this phase. \(E\) is an empirical value in the range of [1,3]. \(C_{\text{tot}}\) represents the total capacitance connected to the common-mode voltage \(V_{cm}\) during the hold phase \(T_\text{h}\).

\subsection{Dependency Graph Construction}
\label{sec:method_dependency_graph}



\begin{figure}[t]
    \centering
    \includegraphics[width=\columnwidth]{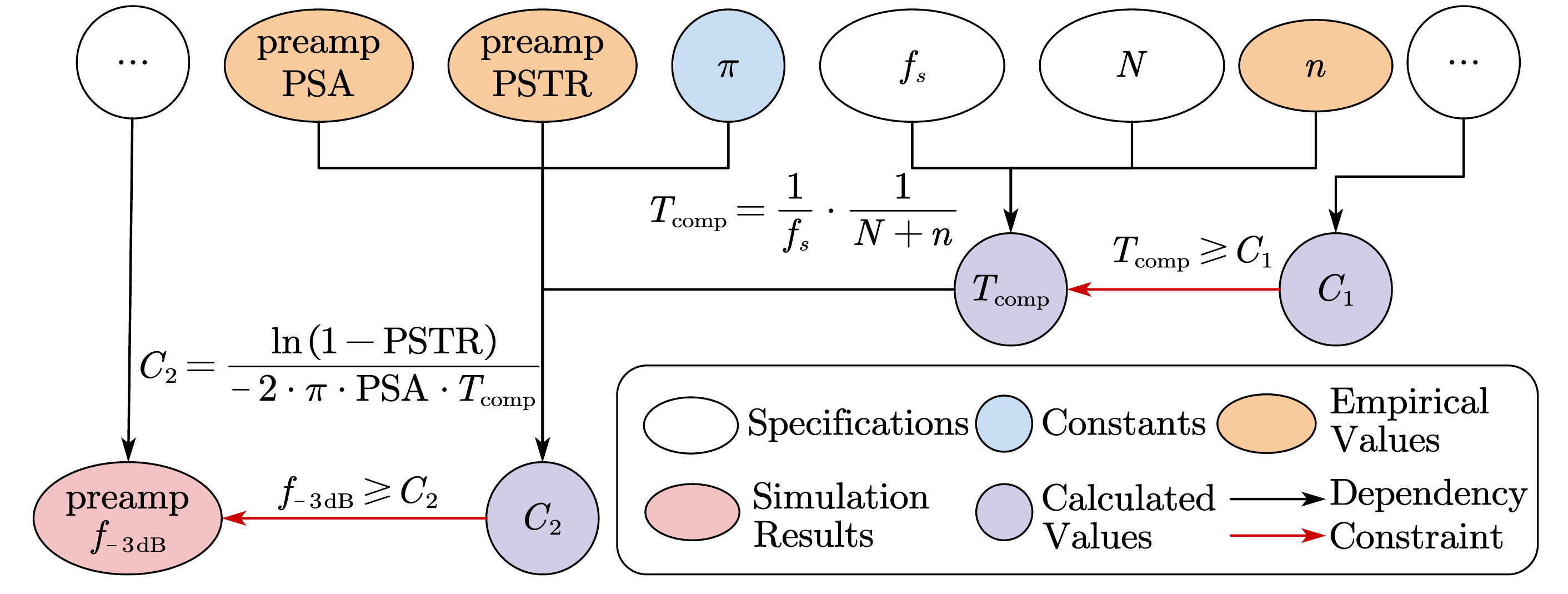}
    \caption{Dependency graph example. Variables and equations are explained in detail in Fig. \ref{fig:Overall_design_flow}}
    \label{fig:dependency_graph}
\end{figure}

Now that we have all the relevant variables and relations, we are ready to use this information for our optimization framework. As shown in Table \ref{tab:variables_and_relations}, each variable may appear on either side of a relation; however, we ensure that only a single variable appears on the left-hand side (LHS). This establishes a \textbf{dependency} from the LHS variable to all variables on the right-hand side (RHS). By considering variables as \textbf{nodes} and dependencies as \textbf{edges}, we construct a \textbf{dependency graph} (see Fig. \ref{fig:dependency_graph} for a partial example). In this work, to simplify the problem, we assume that the dependency graph is cycle-free. Consequently, the graph satisfies the properties of a \textit{directed acyclic graph (DAG)}. This property enables topological sorting, which guarantees that all RHS variables are evaluated before their corresponding LHS expressions are computed or verified.

A subcircuit's performance specifications are derived once all related constraints are evaluated through dependency graph propagation. This effectively partitions the system-level performance specifications to subcircuits. The flow then performs \textit{subcircuit-level} BO on the subcircuit individually to meet the corresponding specifications. After obtaining  simulation results for constraint verification, the flow propagates the performance metrics down the dependency graph in topological order to compute constraints for other subcircuits. For instance, as described in Section~\ref{sec:method_subcircuits}, the comparator offset $V_{os}$ is used to set a constraint on the preamplifier gain $A_V$. In this way, subcircuit-level optimization, simulation, and constraint verification are serialized.

\begin{table*}[htbp]
\caption{Comparison of different works}
\begin{center}
\resizebox{\textwidth}{!}{
\begin{tabular}{|c|cc|cc|ccc|cc|cc|}
\hline  
                     & \multicolumn{2}{c|}{Huang$^{\mathrm{\dagger}}$ \cite{huang2015systematic}}         & \multicolumn{2}{c|}{Ding{$^{\mathrm{\dagger}}$} \cite{ding2018hybrid}}          & \multicolumn{3}{c|}{ Hassanpourghad{$^{\mathrm{*}}$} \cite{hassanpourghadi2021module}}                         & \multicolumn{2}{c|}{Liu{$^{\mathrm{\#}}$} \cite{liu2021opensar}}   & \multicolumn{2}{c|}{Our Work{$^{\mathrm{*}}$}}      \\ \hline
Resolution           & \multicolumn{1}{c|}{10}    & 12    & \multicolumn{1}{c|}{8}     & 12    & \multicolumn{1}{c|}{6}    & \multicolumn{1}{c|}{8}     & 10     & \multicolumn{1}{c|}{10}    & 12    & \multicolumn{1}{c|}{12}    & 14    \\ \hline
Sampling rate (MS/s) & \multicolumn{1}{c|}{50}   & 0.6    & \multicolumn{1}{c|}{32}    & 1     & \multicolumn{1}{c|}{500}   & \multicolumn{1}{c|}{340}   & 200   & \multicolumn{1}{c|}{100}   & 1     & \multicolumn{1}{c|}{1}     & 0.5   \\ \hline
Technology (nm)      & \multicolumn{1}{c|}{90}   & 180    & \multicolumn{2}{c|}{40}            & \multicolumn{3}{c|}{65}                                         & \multicolumn{2}{c|}{40}            & \multicolumn{2}{c|}{28}            \\ \hline
ENOB                 & \multicolumn{1}{c|}{8.4}  & 10.1   & \multicolumn{1}{c|}{7.6}   & 9.9   & \multicolumn{1}{c|}{5.92}  & \multicolumn{1}{c|}{7.75}  & 9.18  & \multicolumn{1}{c|}{9.1}   & 11.1  & \multicolumn{1}{c|}{11.38} & 12.99 \\ \hline
SNDR (dB)             & \multicolumn{1}{c|}{52.1}  & 62.7  & \multicolumn{1}{c|}{47.4}  & 61.1  & \multicolumn{1}{c|}{37.3}  & \multicolumn{1}{c|}{48.4}  & 57.8  & \multicolumn{1}{c|}{56.3}  & 68.8  & \multicolumn{1}{c|}{70.3}  & 79.99 \\ \hline
\(\text{FOM}_S\) (dB)           & \multicolumn{1}{c|}{159.7} & 152.2 & \multicolumn{1}{c|}{156.7} & 165.8 & \multicolumn{1}{c|}{153.2} & \multicolumn{1}{c|}{159.6} & 167 & \multicolumn{1}{c|}{166.8} & 176.0 & \multicolumn{1}{c|}{168.7} & 172.1 \\ \hline
\(\text{FOM}_W\) (fJ/c.step)     & \multicolumn{1}{c|}{26.7}   & 500  & \multicolumn{1}{c|}{30.7}  & 18.4  & \multicolumn{1}{c|}{42.9}  & \multicolumn{1}{c|}{35.5}  & 18.5  & \multicolumn{1}{c|}{10.8}  & 4.3   & \multicolumn{1}{c|}{27.3}  & 37.8  \\ \hline
TOTAL TIME                          & \multicolumn{1}{c|}{-} & - & \multicolumn{1}{c|}{-} & - & \multicolumn{1}{c|}{-} & \multicolumn{1}{c|}{-} & - & \multicolumn{1}{c|}{1h52m42.8s*} & 2h1m27.2s* & \multicolumn{1}{c|}{1h2m44s} & 1h8m31s \\ \hline
\multicolumn{12}{l}{-Unreported value. $^{\mathrm{\dagger}}$Tapeout measurement. $^{\mathrm{\#}}$Simulated results with layout R+C+CC parasitic extraction. $^{\mathrm{*}}$Schematic simulation results.}
\end{tabular}
}
\end{center}
\label{tab:results}
\end{table*}

\subsection{Optimization Procedure}
\label{sec:method_bayesian_optimization}

As shown in Fig. \ref{fig:Overall_design_flow}, BO drives the main optimization efforts in this work. After generating the dependency graph, we employ a \textbf{system-level} optimization loop to obtain a set of empirical values that balance performance and feasibility. The loop terminates when either all constraints are satisfied or the maximum number of iterations is reached. In practice, as demonstrated in Section~\ref{sec:experiments}, all constraints are met within a single iteration of system-level optimization.

The \textbf{subcircuit-level} BO directly tunes device sizing (e.g. transistor length $l$, width $w$) to optimize performance. The allowed sizing range is given by the PDK. 
If any subcircuit performance metric fails to meet its constraint, the deviation is incorporated into the system-level BO loss. Since different performance metrics span several orders of magnitude (e.g., preamplifier bandwidth on the order of $10^{[5, 8]}$ versus comparator propagation delay on the order of $10^{[-10, -9]}$), larger values could overwhelm smaller ones. To normalize these differences, each loss is evaluatedon a logarithmic scale. For instance, if a bandwidth of 100$M$ is desired but only 10$M$ is achieved, the loss incurred is $log(100M) - log(10M) = 1$ instead of $100M - 10M = 90M$. Conveniently, all target values in this experiment are positive, allowing this approach; otherwise, a more advanced normalization method would be required, which is discussed in Section~\ref{sec:conclusion}.

\section{Experimental Results}
\label{sec:experiments}

The entire framework is implemented in Python 3.9, executed on a CentOS 7.9.2009, on an Intel Xeon Gold 5320 CPU with 96 cores. The experiments are performed using TSMC 28nm technology with Cadence's Spectre simulator, which restricts our transistor length, width, and number of fingers to [30nm, 1$\mu$m], [100nm, 3$\mu$m], and [1, 100] respectively. We use Bayesian optimization (BO) implemented by the Optuna~\cite{akiba2019optuna} library, and limit our total simulation count to 5000. Subcircuit-level BO is parallelized at 20 concurrent threads.

Table \ref{tab:results} presents the results given two sets of performance specifications, compared against previous work. Table \ref{tab:runtime} presents the runtime breakdown in terms of real-time seconds and BO iterations. For reference, we also list the total number of parameters tuned in each subcircuit. 

\begin{table}[htbp]
\caption{Runtime breakdown in seconds (BO iterations). Parameter counts are listed next to subcircuit name.}
\label{tab:runtime}
\begin{center}
\begin{tabular}{|c|c|c|c|}
\hline
\textbf{Resolution} & \textbf{12 bit} & \textbf{14 bit} \\ \hline
\textbf{Sampling rate} & \textbf{1MS/s}  & \textbf{500KS/s} \\ \hline \hline
\textbf{Unit Capacitor (2)} & 7 (34) & 11 (52) \\ \hline
\textbf{Bootstrap (22)} & 490 (31) & 1206 (98) \\ \hline
\textbf{Comparator (8)} & 981 (1315) & 20 (31) \\ \hline
\textbf{Pre-amplifier (30)} & 1661 (2147) & 2290 (2848) \\ \hline
\textbf{System-Level (4)} & 3764 (1) & 4111 (1) \\ \hline
\end{tabular}
\end{center}
\end{table}

Overall, the 14-bit 500 KS/s design achieves an SNDR of 80 dB with a power consumption $P$ of 153.6 µW, while the 12-bit 1 MS/s design achieves an SNDR of 70.3 dB with a power consumption of 72.81 µW. We calculate the Schreier and Walden performance metrics (FOM) as follows:
\begin{align}
\text{FOM}_S &= \text{SNDR} + 10 \log \left( \frac{f_s}{2P} \right) \\
\text{FOM}_W &= \frac{P}{f_s\cdot 2^{\text{ENOB}}}
\end{align}
where SNDR is the signal-to-noise-distortion ratio, \(f_s\) is the sampling frequency, ENOB is the effective number of bits for the SAR ADC, and \(P\) is the power consumption. A comparison with previous work is presented in Table \ref{tab:results}.

In the 14-bit 500 KS/s design, 62\% of the power consumption is attributed to the SAR logic. This is due to the higher power consumption associated with the use of synchronous SAR logic. The power consumption might be further optimized if asynchronous SAR was used. In the 12-bit 1 MS/s design, the power consumption of comparator accounts for a larger proportion of the total power consumption, as the dynamic power of the comparator rises in high-speed designs.

In terms of optimization efficiency, as we increase the desired ADC resolution from 12 to 14 and reduce sampling frequency from 1M to 500k, we notice a change in runtime. The comparator requires fewer iterations to meet specifications, while the bootstrap circuit and the pre-amplifier take longer. This is because the higher resolution requirement sets harder specifications for the bootstrap and pre-amplifier circuits. Overall, the proposed framework arrives at a similar FoM and runtime to the previous work, showing the effectiveness of the automated method even without manual intervention.


\section{Discussion and Future Work}
\label{sec:conclusion}

In this work, we propose an autonomous SAR ADC sizing framework. We construct a dependency graph from predefined analytical equations, then use the topological order of the graph to serialize a system-level optimization loop. Within the loop, we partition system-level specifications to each subcircuit and optimize them individually. Finally, the flow verifies the design constraints and decides whether to continue the optimization loop. We verified the proposed framework on two different sets of performance specifications, and successfully produced functional SAR ADCs with comparable performance to previous autonomous sizing frameworks.

For future work, our method will be improved on three different fronts. Firstly, the acyclic dependency assumption will be released, which allows us to support any set of analytical equations. Next, we will employ offline sizing methods, where the subcircuit-level BO are performed from pre-collected simulation data, so that we pay a one-time cost for all sizing, and therefore runtime no longer scales linearly with the number of system-level iterations. This should allow us to tackle less thoroughly collected equations and more aggressive empirical value ranges. Finally, we will expand our method to different types of circuits, as the current methodology is not designed specifically for SAR ADCs and may work for other hierarchical AMS designs as well.

\balance
\bibliographystyle{ieeetr}
\bibliography{refs}



\end{document}